\documentclass[prl,superscriptaddress,twocolumn,showpacs]{revtex4}
\usepackage{amsmath,amssymb,graphicx,color,bm,soul}

\begin{document}

\title{Bistability in Microcavities with Incoherent Optical or Electrical Excitation}

\author{O. Kyriienko}
\affiliation{Science Institute, University of Iceland, Dunhagi-3, IS-107, Reykjavik, Iceland}
\affiliation{Division of Physics and Applied Physics, Nanyang Technological University 637371, Singapore}

\author{T. C. H. Liew}
\affiliation{Division of Physics and Applied Physics, Nanyang Technological University 637371, Singapore}
\affiliation{Mediterranean Institute of Fundamental Physics, 31, via Appia Nuova, Rome, 00040, Italy}

\author{E. A. Ostrovskaya}
\affiliation{Nonlinear Physics Centre, Research School of Physics and Engineering, The Australian National University, Canberra ACT 0200, Australia}

\author{O. A. Egorov}
\affiliation{Institute of Condensed Matter Theory and Solid State Optics, Abbe Center of Photonics, Friedrich-Schiller-Universit\"at Jena, Jena 07743, Germany}

\author{I. A. Shelykh}
\affiliation{Science Institute, University of Iceland, Dunhagi-3, IS-107, Reykjavik, Iceland}
\affiliation{Division of Physics and Applied Physics, Nanyang Technological University 637371, Singapore}

\date{\today}

\begin{abstract}
We consider a quantum well embedded in a zero-dimensional microcavity with a sub-wavelength grated mirror, where the $x$-linearly polarized exciton mode is strongly coupled to the cavity photon, while $y$-polarized excitons remain in the weak coupling regime. Under incoherent optical or electric pumping, we demonstrate polariton bistability associated with parametric scattering processes. Such bistability is useful for constructing  polaritonic devices with optical or electrical incoherent pumping.
\end{abstract}




\maketitle

{\it Introduction.---}Bistability and hysteresis are fundamental properties of resonant optical nonlinear systems~\cite{Gibbs1985} with applications in optical memory elements and transistors. Optical nonlinearity ultimately relies on the coupling of light to electronic degrees of freedom, which mediate effective interactions between photons. With an ever-growing effort devoted to the enhancement of light-matter coupling, high nonlinearity allows bistability to be routinely observed in a variety of systems.

One example is the strong light-matter coupling in high quality factor semiconductor microcavities containing quantum well excitons~\cite{Deveaud2007,Kavokin2007}. The resulting exciton-polaritons are known to exhibit bistability, when driven resonantly and coherently ~\cite{Gippius2004,Baas2004,Whittaker2005}, i.e., with a pumping laser tuned to the energy of the exciton-polariton quasiparticle. This allows a variety of related effects to occur, including driven superfluidity~\cite{Carusotto2004}, the suppression of disorder~\cite{Liew2012} and the formation of various structures -- spin patterns~\cite{Shelykh2008,Sarkar2010,Adrados2010}, solitons~\cite{Larionova2008,Egorov2009,Egorov2010,Sich2011,Egorov2013}, and vortex lattices~\cite{Gorbach2010}. Controlled switching has been observed in multi-mode systems, based on the overlapping of states with different energies and wavevectors~\cite{DeGiorgi2012} or the spin degree of freedom~\cite{Amo2010,Adrados2011,Cerna2013}. In addition, bistability underpins theoretical schemes for electro-optic~\cite{Liew2010} and all-optical circuits~\cite{Espinosa2013,Ballarini2013}.

An important step towards low-threshold polaritonic lasing and electrically controlled integrated polaritonic circuits is condensation of electrically injected polaritons, which was recently observed \cite{Nature_incoherentBEC}. The practical potential of electrically injected devices~\cite{Tsintzos2008,Tsintzos2009,Butte2011,Winkler2013} relies on the compatibility of physical mechanisms controlling bistability with incoherent, non-resonant excitation. However, all of the polaritonic devices based on bistability demonstrated to date require a coherent optical excitation, implying that they must be coupled with a laser light that is resonant or near-resonant with the exciton-polariton state.

\begin{figure}
\includegraphics[width=1.0\linewidth]{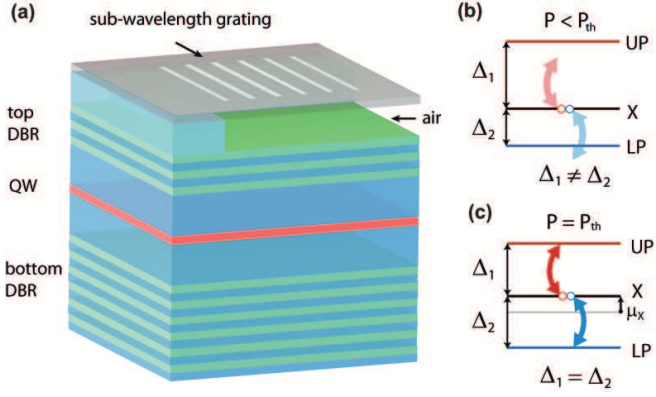}
\caption{(color online). (a) Sketch of the system. The microcavity is formed from distributed Bragg reflectors (DBRs) and a sub-wavelength grating at the top. This allows confinement of the $x$-linearly polarized cavity mode only. (b) and (c): The energy levels corresponding to positive detuning. At small pumping rates of the $\chi_{y}$ exciton parametric processes are not possible, since $\Delta_{1}\neq \Delta_{2}$ (b). Above threshold pumping $P_{th}$, a blueshift in the energy of $\chi_y$ satisfies the parametric scattering condition $\Delta_1 = \Delta_2$ (c).}
\label{fig:sketch}
\end{figure}

In this Letter, we propose a mechanism of optical bistability in a semiconductor microcavity compatible with incoherent non-resonant pumping. We consider the recently developed sub-wavelength grating type microcavity, reported by Zhang \textit{et al.}~\cite{Zhang2013} and illustrated in Fig. \ref{fig:sketch}(a). In such a cavity, light is only confined if it has a specific linear polarization ($x$) such that linearly polarized exciton-polaritons are formed, while a cross-linearly polarized ($y$) exciton mode remains uncoupled to light. We develop evolution equations for occupation numbers and correlators in this system, using a master equation approach~\cite{Shelykh2005} with Lindblad type terms accounting for incoherent pumping and dissipation. First, we consider an optical incoherent excitation on resonance with an exciton mode, which could be polarized in the $y$-direction so as to avoid direct excitation of exciton-polariton modes. Second, we consider the electrical injection of excitons, leading to asymmetric pumping of all three modes. In both cases excitons undergo elastic pair scattering into upper and lower polariton states, in analogy to various intra-branch~\cite{Savvidis2000,Tartakovskii2000,Savasta2005} and inter-branch~\cite{Ciuti2004,Diederichs2006,Xie2012} processes that have been studied in resonantly excited microcavities. This leads to bistable behavior of the system under certain pumping conditions, which we expect to be useful for future polaritonic devices able to work under optical and electrical incoherent pumping.

{\it The model.---}Introducing the field operators of cavity photons, $\hat{\phi}$, and excitons, $\hat{\chi}$, the Hamiltonian of the system can be written as
\begin{align}
\mathcal{\hat{H}}&=E_C\hat{\phi}^\dagger_x\hat{\phi}_x+E_X\left(\hat{\chi}^\dagger_x\hat{\chi}_x+ \hat{\chi}^\dagger_y\hat{\chi}_y\right)+ V\left(\hat{\chi}^\dagger_x\hat{\phi}_x+\hat{\phi}^\dagger_x\hat{\chi}_x\right)\notag\\
&+\alpha_1\left(\hat{\chi}^\dagger_+\hat{\chi}^\dagger_+\hat{\chi}_+\hat{\chi}_+ +\hat{\chi}^\dagger_-\hat{\chi}^\dagger_-\hat{\chi}_-\hat{\chi}_-\right)+2\alpha_2\hat{\chi}^\dagger_+ \hat{\chi}^\dagger_-\hat{\chi}_+\hat{\chi}_-,
\end{align}
where $x$, $y$, $+$ and $-$ subscripts denote linear, cross-linear, circular and cross-circular polarizations, respectively. We consider spin degenerate exciton states with energy $E_X$ and a linearly polarized cavity mode with energy $E_C$. The exciton-photon coupling constant $V$ couples only $x$ polarized states. Non-linear interactions between parallel exciton spins are described by the parameter $\alpha_1$, and interactions between opposite spins are described by $\alpha_2$. The circularly polarized states can be rewritten in terms of linearly polarized states, $\hat{\chi}_+=(\hat{\chi}_x+i\hat{\chi}_y)/\sqrt{2}$ and $\hat{\chi}_-=(\hat{\chi}_x-i\hat{\chi}_y)/\sqrt{2}$. Introducing lower and upper $x$ polarized polariton states, $\hat{\phi}_x=C\hat{\psi}_L-X\hat{\psi}_U$ and $\hat{\chi}_x=X\hat{\psi}_L+C\hat{\psi}_U$, we can diagonalize the linear part of the Hamiltonian:
\begin{align}
\hat{\mathcal{H}}&=E_L\hat{\psi}_L^\dagger\hat{\psi}_L+E_U\hat{\psi}_U^\dagger\hat{\psi}_{U}+\frac{(\alpha_1+\alpha_2)}{2}\left[X^4\hat{\psi}_{L}^\dagger\hat{\psi}_{L}^\dagger\hat{\psi}_{L}\hat{\psi}_{L}\right.\notag\\
&\left.+C^4\hat{\psi}_{U}^\dagger\hat{\psi}_{U}^\dagger\hat{\psi}_{U}\hat{\psi}_{U}+4X^2C^2\hat{\psi}_{L}^\dagger\hat{\psi}_{U}^\dagger\hat{\psi}_{L}\hat{\psi}_{U}+\hat{\chi}^\dagger_y\hat{\chi}^\dagger_y\hat{\chi}_y\hat{\chi}_y\right]\notag\\
&-(\alpha_1-\alpha_2)XC\left(\hat{\psi}_{L}^\dagger\hat{\psi}_{U}^\dagger\hat{\chi}_y\hat{\chi}_y+\hat{\chi}_y^\dagger\hat{\chi}_y^\dagger\hat{\psi}_{L}\hat{\psi}_{U}\right)\notag\\ &+2\alpha_1\left(X^2\hat{\psi}_{L}^\dagger\hat{\psi}_{L}+ C^2\hat{\psi}_{U}^\dagger\hat{\psi}_{U}\right)\hat{\chi}_y^\dagger\hat{\chi}_y+ E_X \hat{\chi}^\dagger_y\hat{\chi}_y.
\end{align}
The polariton energies are given by $E_{U,L}=(E_C+E_{X})/2 \pm \sqrt{\delta^2 + 4V^2}/2$,
where $\delta= E_{C}-E_{X}$ denotes a detuning between cavity photon and exciton modes.
The corresponding Hopfield coefficients $X$ and $C$ can be defined using relations $\{X^2,C^2 \}=(1 \pm \delta/\sqrt{\delta^2+4V^2})/2$.

Given the Hamiltonian, the system evolution follows the master equation for the density matrix $\rho$:
\begin{align}
&\frac{d\rho}{d t}=-\frac{i}{\hbar}\left[\hat{\mathcal{H}},\mathbf{\rho}\right]+ \sum_{a=\chi,\psi_U,\psi_L} P_a \Big(\hat{a}\mathbf{\rho}\hat{a}^\dagger +\hat{a}^\dagger\mathbf{\rho}\hat{a}- \hat{a}^\dagger\hat{a}\mathbf{\rho} \notag\\
& -\mathbf{\rho}\hat{a}\hat{a}^\dagger\Big) +\sum_{a=\chi,\psi_U,\psi_L}\frac{\Gamma_{a}}{2\hbar}\left(2\hat{a}\mathbf{\rho}\hat{a}^\dagger-\hat{a}^\dagger\hat{a}\mathbf{\rho}- \mathbf{\rho}\hat{a}^\dagger\hat{a}\right),
\label{eq:master}
\end{align}
%
where we defined $\hat{\chi}\equiv \hat{\chi}_y$, $P_i$ ($i=X,L,U$) are the rates of incoherent pumping for different modes, and we accounted for different decay rates $\Gamma_{X}$, $\Gamma_{U}= X^{2}\Gamma_{C}+C^{2}\Gamma_{X}$, and $\Gamma_{L}= C^{2}\Gamma_{C}+X^{2}\Gamma_{X}$, corresponding to $\chi$, $\psi_{U}$ and $\psi_{L}$ modes. $\Gamma_{C}$ denotes the cavity photon linewidth. We will consider two cases: 1) Incoherent optical polarized pump ($P_X > 0,~ P_{L,U} = 0$), where only $\chi_y$ excitons are pumped; 2) Electrical pump ($P_{X,L,U} > 0$), where all three modes are asymmetrically populated. The density matrix allows the calculation of observable quantities as $\langle\hat{\mathcal{O}}\rangle= \mathrm{tr}\{\hat{\mathcal{O}}\rho\}$. From Eq.~(\ref{eq:master}) we can thus derive a set of equations for the population numbers of polaritons $N_{L,U} = \langle\hat{\psi}_{L,U}^\dagger\hat{\psi}_{L,U}\rangle$ and excitons $N_{X}=\langle\hat{\chi}^\dagger\hat{\chi}\rangle$ \cite{Supplemental}:
\begin{align}
\frac{dN_{L}}{dt}&=-\frac{2\alpha_-}{\hbar}XC\Im m\left\{A\right\}-\frac{\Gamma_{L}}{\hbar}N_{L}+ P_L,\label{eq:dNLdt}\\
\frac{dN_{U}}{dt}&=-\frac{2\alpha_-}{\hbar}XC\Im m\left\{A\right\}-\frac{\Gamma_{U}}{\hbar}N_{U}+ P_U,\label{eq:dNUdt}\\
\frac{dN_{X}}{dt}&=\frac{4\alpha_-}{\hbar}XC\Im m\left\{A\right\}-\frac{\Gamma_{X}}{\hbar}N_X + P_X,\label{eq:dNXdt}
\end{align}
where the correlator $A=\left<\hat{\psi}_L^\dagger\hat{\psi}_U^\dagger\hat{\chi}\hat{\chi}\right>$ and we used bosonic commutation rules to evaluate the commutators in Eq.~(\ref{eq:master}). Here we introduced the shorthand notation $\alpha_{\pm}=\alpha_1 \pm \alpha_2$. Similarly, one can derive an evolution equation for $A$, truncating third order correlators as products of lower order correlators~\cite{Shelykh2005}, \textit{e. g.}, $\left<\hat{\psi}_L^\dagger\hat{\psi}_L^\dagger\hat{\psi}_L\hat{\psi}_U^\dagger\hat{\chi}\hat{\chi}\right>\approx \left<\hat{\psi}_L^\dagger\hat{\psi}_U^\dagger\hat{\chi}\hat{\chi}\right> \left<\hat{\psi}_L^\dagger\hat{\psi}_L\right> = A N_{L}$. This yields~\cite{Supplemental}:
\begin{align}
&\frac{dA}{dt}=\frac{i}{\hbar}(\beta_1 A + \beta_2)-\frac{\Gamma}{\hbar}A,
\label{eq:dAdt}
\end{align}
where $\Gamma = (\Gamma_{L}+\Gamma_{U})/2 + \Gamma_{X}$, and we defined the auxiliary functions
\begin{align}
&\beta_1 = \delta +  \alpha_+( 2X^2C^2-1)  \\
& -2(\alpha_+ + \alpha_-) ( X^2N_L + C^2N_U)-(\alpha_+-\alpha_ -)N_X \notag\\
& + \alpha_+ [N_L(X^4+2X^2C^2) +N_U(C^4+2X^2C^2)] \notag\\
&\beta_2 = \alpha_- XC [  2N_L N_U (2N_X+1) - (N_L+N_U+1)N_X^2 ].\notag
\end{align}
Steady state solutions can be found by setting $dN_{U,L,X}/dt=0$ and $dA/dt=0$. Separating real and imaginary parts of $A$, one obtains $\Im m\{A\}=\beta_{2}\Gamma/(\beta_{1}^2 + \Gamma^2)$. Then, the system of equations for steady states can be reduced to the  single third-order equation for $N_L$:
\begin{equation}
\label{eq:NL}
N_{L}\Gamma_{L}(\beta_{1}^2 + \Gamma^2)+ 2 \alpha_- XC\beta_{2}\Gamma - \hbar P_L (\beta_{1}^2 + \Gamma^2) = 0,
\end{equation}
where $\beta_1$ and $\beta_{2}$ can be re-written in terms of $N_L$ using
\begin{align}
\label{eq:NU}
&N_{U} = N_{L} \frac{\Gamma_{L}}{\Gamma_{U}} + \frac{\hbar (P_U - P_L)}{\Gamma_{U}},\\
\label{eq:NX}
&N_{X} = -2N_{L} \frac{\Gamma_{L}}{\Gamma_{X}} + \frac{\hbar (P_X + 2 P_L)}{\Gamma_{X}}.
\end{align}
We note that the characteristic energy scale of the system is defined by Rabi splitting $\Omega_{R} \equiv 2V$. In particular, the important parameter which governs the behavior of the system is the ratio of detuning to Rabi energy, $\delta/2 V$.

Since $\beta_2$ has a cubic dependence on occupation numbers, there are in general three solutions for each equation. Here we shall be interested only in real and non-negative solutions, which for certain parameters can represent bistable behavior of the system. While analytical solution of Eq. (\ref{eq:NL}) is possible, the resulted expression is too bulky to be presented here. Further we restrict ourselves to numerical treatment of the problem only.

It is also important to note that we restrict our treatment to a micropillar structure of small radius, which can be considered as a zero-dimensional photonic structure. This allows us to disregard spatial dynamics in the system, leading to the simplified three level scheme described above and shown in Fig.  \ref{fig:sketch}(b). Consequently, exact frequency matching is required to start populating polariton modes. However, for larger structures where states with non-zero wave-vectors appear, the parametric scattering conditions can change, with subsequent modification of $N_{L,U,X}$ solutions.

{\it Incoherent bistability with polarized optical pump.---}First we consider excitation with a very broad LED beam, representing an incoherent source that is linearly polarized in the $y$ direction using a polarizer. This pumps only the weakly coupled excitonic mode $\chi_{y}$ ($P_{L,U} = 0$). The system depends strongly on the mode detuning $\delta = E_{C}-E_{X}$. For positive detuning and weak pump, the condition of parametric scattering between exciton and polariton modes is not satisfied [Fig. \ref{fig:sketch}(b)]. However, the increase of exciton concentration with $P_X$ leads to a blueshift of the mode, $\mu_{X}$, and at a threshold intensity, which can be estimated as $P_{th}=\delta \Gamma_{X} / 2 \hbar \alpha_+$, the energy distance between modes becomes equal and parametric scattering is possible [Fig. \ref{fig:sketch}(c)]. This conversion of two $y$-polarized excitons into upper and lower polariton starts to populate otherwise empty polariton modes.
\begin{figure}
\centering
\includegraphics[width=1.0\linewidth]{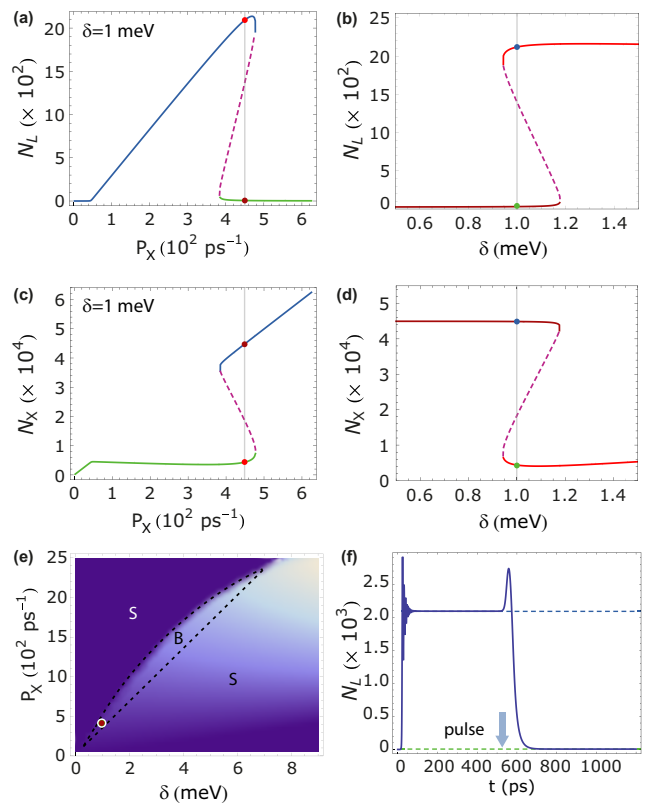}
\caption{(color online). Steady-state solutions in a sub-wavelength grating microcavity. (a) Dependence of the polariton population for varying pump power (with fixed $\delta =1$ meV). The vertical gray line shows a selected pump power at which switching can be observed between the two stable states marked by spots. (b) Dependence of the polariton population for varying $\delta$ [the pump power is the same as that marked by the vertical gray line in (a)]. (c, d) Variation of the exciton densities, corresponding to (a) and (b). (e) Phase diagram of the system in the $P_X/\delta$ coordinate plane. The intensity of the plot corresponds to the population of the highest available polariton state. The dashed curve denotes the region B in which multiple states are possible, and label S refers to the region of parameters corresponding to a single solution. (f) Time dynamic simulation for the green point in (e). The addition of a pulse to the \emph{cw} pump at $t=550$ ps induces switching between the high intensity and low intensity states marked in (a) and (b).}
\label{fig:ResultSWG}
\end{figure}
%


The results of calculations are shown in Fig.~\ref{fig:ResultSWG} for experimentally reported parameters~\cite{Zhang2013}: $V=6$ meV; $\Gamma_{C}=\hbar/\tau_{C}$; where $\tau_{C}=5$ ps is the cavity photon lifetime; $\tau_{X}=100$ ps \cite{Sermage}; and $S=25~\mu$m$^2$ is the sample area. The exciton-exciton scattering strength in the triplet channel was calculated as $\alpha_1=6E_Ba_B^2/S$~\cite{Tassone1999}, where $E_B=4.8$ meV and $a_B=11.6$ nm are the exciton binding energy and Bohr radius in GaAs, respectively. The singlet interaction constant was shown to vary in a broad range with varying detuning~\cite{Vladimirova2010}, or can be tuned by other means \cite{PLMCN14}. In the following calculations we use the value $\alpha_2=0.4\alpha_1$ reported in Ref. \cite{Paraiso2010}.

For a fixed detuning, Fig.~\ref{fig:ResultSWG}(a) shows that at very low pump intensity parametric scattering does not take place such that the polariton population is zero. With increasing intensity parametric scattering is turned on, increasing the polariton population at the expense of limiting the exciton population [Fig.~\ref{fig:ResultSWG}(c)]. At high pump powers, the unequal blueshifts of the modes in the system causes the switching off of the parametric scattering, with the polariton population returning to small values close to zero. A similar effect was seen in Ref.~\cite{Whittaker2005} in resonantly excited parametric oscillators. Here the lower intensity branch has vanishingly small occupation comparing to higher branch or excitonic mode occupancies. While being typically less than unity for small detuning, it can reach order of one in the higher detuning region. This may not be sufficient for parametric scattering as the occupation of $N_U$ may be suppressed at the same detuning. However, one can still expect energy relaxation mechanisms involving higher energy states \cite{Porras2002}, which can potentially lead to polariton lasing in the micropillar structure \cite{Bajoni2008}. For a clear demonstration of bistability, we choose parameters where parametric scattering is dominant or the occupancy of $N_L$ is too small for polariton lasing.

Here, there is a concurrent bistability present in the system shown, for example, by the vertical gray line at which the system can exist in a high or zero polariton intensity state. Note that while the excitonic mode shows conventional S-shape bistability, the shape of lower polariton bistability is different. This can be easily explained keeping in mind relations between modes given by Eq. (\ref{eq:NX}). Additionally, the switching between high and low intensity states can be controlled with detuning between modes, as shown in Figs.~\ref{fig:ResultSWG}(b) and (d).

The system is further characterized by the phase diagram in the $P_X/\delta$ plane shown in Fig.~\ref{fig:ResultSWG}(e). The dashed boundary marks the region in which multiple solutions are present. Taking parameters at the green dot in Fig. \ref{fig:ResultSWG}(e), switching between stable states can be demonstrated by numerical solution of Eqs.~(\ref{eq:dNLdt})--(\ref{eq:dAdt}), as shown in Fig.~\ref{fig:ResultSWG}(f). Assuming that the system is initially unoccupied and switching on the continuous wave (\emph{cw}) pump, the system quickly stabilizes into the higher polariton density state shown in Figs.~\ref{fig:ResultSWG}(a)-(d). A short pulse applied at $t=550$ ps is able to switch the system into the lower density state. During the intervals without pulses, the system is stable, remaining in the state set by its history.
We note that the appearance of the bistability window also depends on the singlet interaction constant $\alpha_2$, favouring small $\alpha_- = \alpha_1 - \alpha_2$. However, the bistable behavior can still be observed for small and even zero $\alpha_2$ going to higher detuning region.
\begin{figure}
\includegraphics[width=1.0\linewidth]{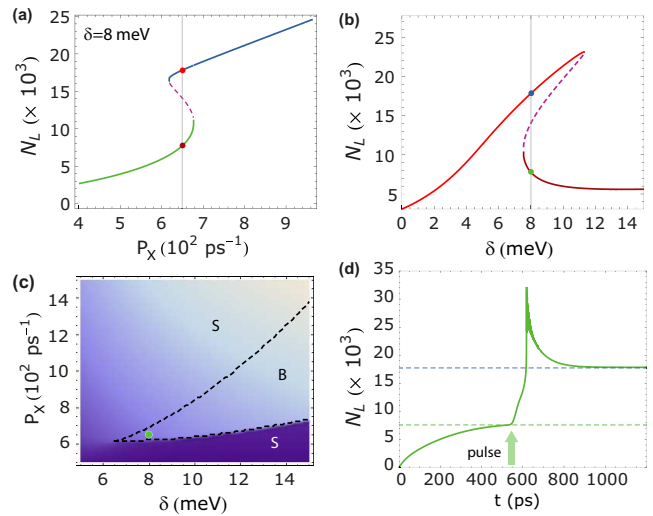}
\caption{(color online). Steady-state solutions in a sub-wavelength grating microcavity with electrical pumping. (a) Dependence of the lower polariton $N_{L}$ population on pumping strength $P_{X}$ plotted for detuning $\delta = 8$ meV. (b) Lower polariton population shown as a function of detuning for fixed pump $P_{X}=6.5\times 10^2$ ps$^{-1}$. (c) Phase diagram plotted in $P_{X}/\delta$ marking bistable (B) and single solution (S) regions. (d) Simulation of the lower polariton occupation number as a function of time, showing switching between low and high intensity states (dashed lines). An additional pulse arrives at $t=550$ ps. Parameters are shown by the green dot in (c).}
\label{fig:el}
\end{figure}

{\it Electrically pumped bistability.---}Next we consider the case of electrical pumping, where both $x$- and $y$-polarized excitonic modes are pumped ($P_{L,U} \neq 0$). The dependence of the lower polariton occupation number $N_{L}$ as a function of pumping strength $P_X$ is shown in Fig.~\ref{fig:el}(a) and reveals electrically pumped bistability. The key difference between electrical pumping and the previously considered optical polarized pump is an additional excitation of polariton modes with rates $P_L$ for lower polaritons and $P_U$ for upper polaritons, which in general are different in the case of non-zero detuning and due to presence of thermalization processes. Particularly, the upper polariton mode occupation is typically small, while relaxation mechanisms lead to effective incoherent pumping of lower polariton modes. This leads to an asymmetric blue-shift of both exciton and polariton energies, and consequent shift of $P_{th}$ to higher pumping rates. Additionally, an incoherent bistability with electrical injection requires larger positive detuning than for the optical case and negative singlet interaction constant $\alpha_2$ [see Fig. \ref{fig:el}(b)]. However, these values of detuning are easily accessible in current experiments \cite{FengLi2013,Trichet2013}.

The phase diagram in the $P_{X}/\delta$ plane is shown in Fig.~\ref{fig:el}(c). Consequently, the width of the pumping strength dependent bistability window diminishes compared to the optical case, although is still fully accessible. Finally, the switching behaviour of the modes, calculated similarly to the optical pump case, confirms stability of the bistable solutions [Fig. \ref{fig:el}(d)]. For the calculations in Fig. \ref{fig:el} we considered the system with single GaAs QW in the cavity with light-matter interaction constant being $V= 2$ meV, $\alpha_2= -0.4 \alpha_1$, and effective pumping amplitudes of polariton states are $P_U=0.01 C^2 P_X$ and $P_L = 0.1 X^2 P_X$.

{\it Conclusion.---}To summarize, we considered parametric scattering between $y$-polarized excitons and $x$-polarized polaritons in a sub-wavelength grating microcavity. We demonstrated bistable behaviour in the cases of both incoherent optical polarized pumping and electrical exciton excitation. This theoretical evidence for fast-response off-resonantly driven bistability lays a foundation for an incoherent analogue of driven polaritonic circuits~\cite{Espinosa2013}, allowing for hybrid electro-optic designs.

The work was supported by FP7 IRSES project POLAPHEN and Tier 1 project ``Polaritons for novel device applications''. O. K. acknowledges the support from the Eimskip Fund.

\end{document}